\documentclass[onecolumn,shortnote]{jpsj3}
\usepackage{color}
\usepackage{ulem}

\title{Breakdown of Chemical Scaling for Pt-Doped CaFe$_2$As$_2$}

\author{
\name{Kazutaka \surname{Kudo}}$^{1,2}$\thanks{E-mail address: kudo@science.okayama-u.ac.jp}, 
\name{Masakazu \surname{Kobayashi}}$^{1,2}$, 
\name{Satomi \surname{Kakiya}}$^{1,2}$, 
\name{Masataka \surname{Danura}}$^{1,2}$, 
and \name{Minoru \surname{Nohara}}$^{1,2}$
}

\inst{
$^{1}$Department of Physics, Okayama University, Okayama 700-8530, Japan \\
$^{2}$JST, Transformative Research-Project on Iron Pnictides (TRIP), Chiyoda, Tokyo 102-0075, Japan \\
}

\kword{iron-based superconductors, phase diagram, CaFe$_2$As$_2$, Ca-Fe-Pt-As, Pt doping}

\begin{document}
\maketitle

The electronic phase diagram of Co-doped CaFe$_2$As$_2$ captures the generic features of iron-based superconductors\cite{rf:Ishida,rf:Paglione,rf:Johnston}. 
The parent compound CaFe$_2$As$_2$ exhibits a phase transition from a paramagnetic metal (PM) phase to an antiferromagnetic metal (AFM) one upon cooling at a transition temperature $T_{\rm N}$ = 170 K, as shown in Fig. 1(a)\cite{rf:Harnagea}. 
Almost simultaneously, system exhibits a structural phase transition from a tetragonal phase to an orthorhombic one at $T_{\rm s}$\cite{rf:Harnagea}. 
The partial chemical substitution of Co for Fe suppresses the AFM and orthorhombic phase, and a superconducting phase appears. 
The maximum superconducting transition temperature $T_{\rm c} =$ 20 K is observed near the critical concentration of Co, i.e., $x$ $\simeq$ 0.06, at which the AFM ordering is completely suppressed in Ca(Fe$_{1-x}$Co$_x$)$_2$As$_2$\cite{rf:Harnagea}. 
The superconducting phase disappears on further doping. 
Ni-doped CaFe$_2$As$_2$ shows similar phase diagram\cite{rf:Kumar}; however, the critical concentration of Ni, at which the AFM phase is suppressed and superconducting phase appears, is almost half of that for Co-doped CaFe$_2$As$_2$, as shown in Fig.~1(a). 
The above-mentioned observations imply that the dependence of $T_{\rm N}$, $T_{\rm s}$, and $T_{\rm c}$ on the doping level $x$ can be interpreted in terms of the difference in the number of valence electrons between the doped transition metal (TM) and iron \cite{rf:Saha,rf:Ni}, namely, the chemical scaling of the electronic phase diagram. 
Such scaling has been reported for SrFe$_2$As$_2$ as well as BaFe$_2$As$_2$ with different TMs\cite{rf:Saha,rf:Ni}.

In this paper,  we report a breakdown of the scaling of $T_{\rm N}$ on chemical doping $x$ for Pt-doped CaFe$_2$As$_2$. 
We demonstrate that the AFM phase persists until the Pt content $x$ reaches its solubility limit at 0.08. 
This behavior is contradictory to that of Ni-doped CaFe$_2$As$_2$, in which the AFM is suppressed at $x$ = 0.03, although both Ni and Pt are isovalent.
Thus, we observe that Ca(Fe$_{1-x}$Pt$_x$)$_2$As$_2$ does not exhibit superconductivity.

\begin{figure}[t]
\begin{center}
\includegraphics[width=8cm]{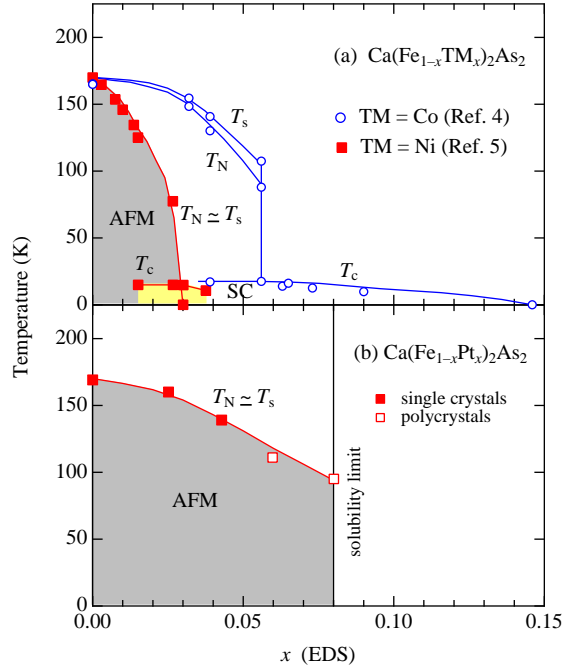}
\caption{
(Color online) (a) Electronic phase diagrams of Ca(Fe$_{1-x}$Co$_x$)$_2$As$_2$ \cite{rf:Harnagea} and Ca(Fe$_{1-x}$Ni$_x$)$_2$As$_2$. \cite{rf:Kumar}
AFM and SC denote antiferromagnetic and superconducting phases, respectively. 
$T_{\rm N}$, $T_{\rm S}$, and $T_{\rm c}$ denote the AFM transition temperature, structural phase transition temperature, and superconducting transition temperature, respectively.
(b) Electronic phase diagram of Ca(Fe$_{1-x}$Pt$_x$)$_2$As$_2$. 
The closed and open squares indicate $T_{\rm N}$ determined from magnetization measurements using single crystals and polycrystals, respectively. 
}
\end{center}
\end{figure}

\begin{figure}[h]
\begin{center}
\includegraphics[width=8cm]{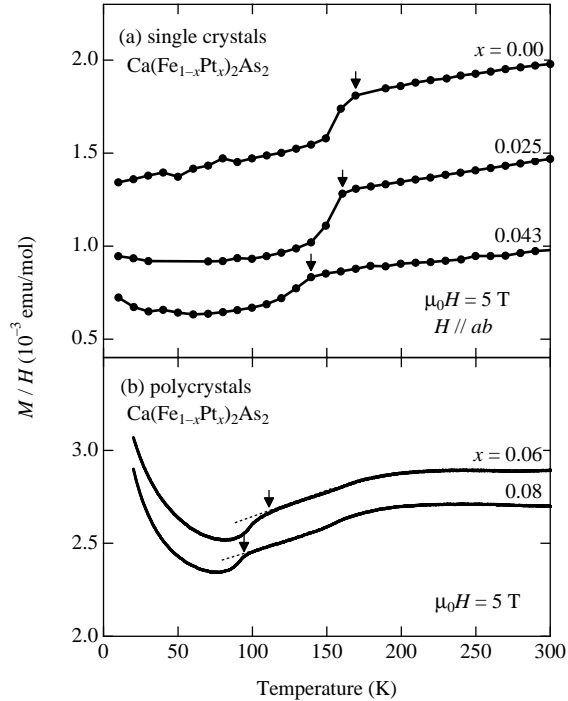}
\caption{Temperature dependence of magnetization divided by magnetic field, $M/H$, for (a) single crystals and (b) polycrystals of Ca(Fe$_{1-x}$Pt$_x$)$_2$As$_2$ in a magnetic field of 5 T. 
The arrows indicate the antiferromagnetic transition temperature $T_{\rm N}$. 
For clarity, $M/H$ is shifted by +0.8 $\times$ 10$^{-3}$ emu/mol and +0.4 $\times$ 10$^{-3}$ emu/mol for $x$ = 0.00 and 0.025, respectively. Broken lines in (b) are guides to the eye. 
}
\end{center}
\end{figure}

Single crystals of Ca(Fe$_{1-x}$Pt$_x$)$_2$As$_2$ ($x$ $=$ 0, 0.025, and 0.043) were grown using a self-flux method. 
Details of this method are given in Ref.~\citen{rf:Danura}. 
Polycrystalline samples of Ca(Fe$_{1-x}$Pt$_x$)$_2$As$_2$ ($x$ $=$ 0.06 and 0.08) were synthesized by a solid-state reaction\cite{rf:Nishikubo}. 
Prescribed amounts of Ca, FeAs, Pt, and As powders or grains were mixed and ground. 
The resulting powder was heated in an evacuated quartz tube at 700 $^\circ$C for 3 h and then at 1000 $^\circ$C for 72 h. 
The obtained samples were characterized by powder X-ray diffraction (XRD) and confirmed to be a single phase of Ca(Fe$_{1-x}$Pt$_x$)$_2$As$_2$.
The Pt content $x$ was determined by energy dispersive X-ray spectrometry (EDS); we used this value of $x$ in the study. 
The maximum value of $x$ obtained by the self-flux growth was approximately 0.043. 
The range of  $x$ could be extended up to 0.08 through a solid-state reaction. 
A single-phase sample with $x$ greater than 0.08 was hardly obtained, indicating that the solubility limit of Pt was at $x$ = 0.08 in Ca(Fe$_{1-x}$Pt$_x$)$_2$As$_2$.

We measured magnetization $M$ using a SQUID magnetometer (Quantum Design). 
The magnetic susceptibility $M/H$ of single crystals exhibited a characteristic $T$-linear behavior in the paramagnetic phase at high temperatures, indicative of magnetic fluctuations \cite{rf:Danura,rf:Zhang}, and subsequently, $M/H$ decreased rapidly at the antiferromagnetic transition temperature $T_{\rm N}$\cite{rf:Danura}, indicated by arrows in Fig.~2(a). 
We expect that a structural phase transition occurs at $T_{\rm s} \simeq T_{\rm N}$, although $T_{\rm s}$ and $T_{\rm N}$ were hardly resolved from the $M/H$ data\cite{rf:Pratt}. 
A similar behavior of $M/H$ was observed in the polycrystalline samples, as shown in Fig. 2(b), although these samples show a tiny Curie tail at low temperatures, that is superposed with an almost temperature independent background (of approximately 2 $\times$ 10$^{-3}$ emu/mol), most probably due to tiny impurities in them. 
No sign of superconductivity is observed in the low-field magnetization (not shown). 
The electronic phase diagram of Ca(Fe$_{1-x}$Pt$_x$)$_2$As$_2$ is obtained on the basis of this data, as shown in Fig.~1(b). 
The antiferromagnetic transition temperature $T_{\rm N}$ (as well as $T_{\rm s}$) decreases with increasing $x$, as is generally observed in iron-based superconductors upon chemical doping. 
However, $T_{\rm N}$ (and $T_{\rm s}$) decreases at a slow rate, and the AFM phase remains intact until $x$ attains the solubility limit at 0.08. 
Thus, superconductivity does not emerge in Ca(Fe$_{1-x}$Pt$_x$)$_2$As$_2$.

The difference in phase diagrams between Ni and Pt doped CaFe$_2$As$_2$ is striking. 
Ni and Pt are isovalent, i.e., both have the same number of valence electrons. 
Therefore, it would be obvious to expect that Ni and Pt will dope almost the same amount of carriers, and thus, $T_{\rm N}$ (and $T_{\rm s}$) will decrease with Ni and Pt doping at an almost same rate; therefore, the chemical scaling will hold. 
Indeed, in the case of SrFe$_2$As$_2$, the AFM phase is suppressed at approximately the same doping level $x$ = 0.07, for both Ni~ \cite{rf:Saha} and Pt. \cite{rf:Nishikubo,rf:Kirshenbaum}
At present, the question why the scaling breaks down in Pt-doped CaFe$_2$As$_2$ is, however, unanswered. 
In order to answer this question, further studies need to be conducted from the viewpoints of structural parameters\cite{rf:Drotziger}, disorder\cite{rf:Vavilov}, magnetic dilution\cite{rf:Dhaka}, and first principles\cite{rf:Wadati,rf:Nakamura}.

Finally, we note that a novel phase, $\beta$-Ca$_{10}$(Pt$_3$As$_8$)(Fe$_{2-x}$Pt$_x$As$_2$)$_5$ ($x$ $=$ 0.16), appears along with Ca(Fe$_{1-x}$Pt$_x$)$_2$As$_2$ ($x$ $=$ 0.08), when we intend to substitute Pt beyond the solubility limit at $x$ = 0.08. 
Interestingly, the former exhibits superconductivity at $T_{\rm c}$ $=$ 13 K\cite{rf:Kakiya,rf:Ni3,rf:Johrendt}, whereas the latter does not, even though the Pt content of the Fe site is almost the same in both (8\%). 
We expect the additional charge carriers (electrons) to be self-doped from the Pt$_3$As$_8$ layers to the FeAs layers of $\beta$-Ca$_{10}$(Pt$_3$As$_8$)(Fe$_{2-x}$Pt$_x$As$_2$)$_5$, although the compound is still in the underdoped region according to a Hall measurement\cite{rf:Kakiya}. 
Further attempt to increase Pt concentration yields $\alpha$-Ca$_{10}$(Pt$_4$As$_8$)(Fe$_{2-x}$Pt$_x$As$_2$)$_5$ ($x$ = 0.36), which exhibits superconductivity at a higher temperature of $T_{\rm c}$ = 38 K\cite{rf:Kakiya}. 
The requirement of such a heavy doping of Pt to achieve superconductivity in $\alpha$-Ca$_{10}$(Pt$_4$As$_8$)(Fe$_{2-x}$Pt$_x$As$_2$)$_5$ is consistent with the inefficiency of Pt in reducing the AFM phase of CaFe$_2$As$_2$.

In conclusion, we found that the substitution of Pt is ineffective in the reduction of AFM ordering as well as in inducing superconductivity in Ca(Fe$_{1-x}$Pt$_x$)$_2$As$_2$, and the chemical scaling of the electronic phase diagram breaks down. 
The Pt-doped CaFe$_2$As$_2$ that does not exhibit superconductivity will, however, provide us an ideal opportunity to elucidate the role of chemical doping in the occurrence of superconductivity in iron-based materials.

\section*{Acknowledgement} 
We thank Prof. T. Kambe for providing technical assistance in magnetization measurements. 
Part of this work was performed at the Advanced Science Research Center, Okayama University. 
This work was partially supported by Grants-in-Aid for Scientific Research from the Japan Society of the Promotion of Science and the Ministry of Education, Culture, Sports, Science and Technology, Japan.

\providecommand{\noopsort}[1]{}\providecommand{\singleletter}[1]{#1}%

\end{document}